\documentclass[twocolumn,showpacs,floatfix,pre,superscriptaddress]{revtex4-1}
\RequirePackage{lineno}

\usepackage{graphicx} 
\usepackage{dcolumn} 
\usepackage{epsf} 
\usepackage{amsmath}
\usepackage{bm} 
\usepackage{setspace}
\usepackage{color}
\usepackage[toc,page]{appendix}

\newcommand{\Gj}[6]{ \begin{matrix}
   #1 & #2 & #3 \\
   #4 & #5 & #6 
  \end{matrix}}

\begin{document}

\title{Green-Kubo stress correlation function at the atomic scale and a long-range bond-orientational ordering in a model liquid}

\author{V.A.~Levashov}

\affiliation{Technological Design Institute of Scientific Instrument Engineering,
Novosibirsk, 630055, Russia}
\affiliation{Landau Institute for Theoretical Physics, Russian Academy of Sciences, 
142432 Chernogolovka, Moscow Region, Russia}


\begin{abstract}
Recently there have been several considerations by different authors of viscosity and the Green-Kubo
stress correlation function from the microscopic perspective. 
In most of these and earlier works the atomic level stress is the minimal element of stress.
It is also possible to consider, for pairwise interaction potentials, 
as the minimal elements of stress, the stress tensors associated with the pairs of interacting particles.
From this perspective, the atomic level stress is not the minimal stress element, but a sum of
all pair stress elements in which involved a selected particle.
In this paper, we consider the Green-Kubo stress correlation function from a microscopic perspective
using the stress tensors of interacting pairs as the basic stress elements.
The obtained results show the presence of a long-range bond-orientational order in the studied model liquid and naturally
elucidate the connection of the bond-orientational order with viscosity. 
It turns out that the long-range bond-orientational order is more clearly expressed in the pairs' stress
correlation function than in the atomic stress correlation function. 
On the other hand, previously observed stress waves are much better expressed 
in the atomic stress correlation functions.
We also address the close connection of our approach with the previous bond-orientational order considerations. 
Finally, we consider the probability distributions for the bond-stress and atomic stress correlation 
products at selected distances. 
The character of the obtained probability distributions raises questions about 
the meaning of the average correlation functions at large distances.
\end{abstract}



\today
\maketitle


\section{Introduction}\label{s:intro}

It is generally believed that understanding 
the behavior of supercooled liquids, 
the phenomenon of the glass transition, 
and the behavior of amorphous solids
requires an understanding of the structure and dynamics of 
these materials at the atomic 
scale \cite{HansenJP20061,Boon19911,HarrowellP2012,Wang2012,Williams2015,Pastore2017,EMa20111,ChenYQ2013}.

One of the directions, in discussing the liquids and amorphous materials from the microscopic perspective,
concerns the considerations of stresses in these materials at the atomic scale.
Since the beginning of computer age, the stress states of model supercooled liquids and
amorphous solids have been addressed at the atomic scale by many researchers 
\cite{Kirkwood19501,Egami19801,Egami19802,Egami19821,Aur19871,Chen19881,Hoheisel19881,Woodcock19911,Woodcock20061,
Goldhirsch2002,Kust2003a,Kust2003b,Picard20041,Leonforte20041,Tanguy20061,Tsamados20081,Leporini20121,HarrowellP20121,ChenYQ2013,
Ashwin20151,IlgP20151,HarrowellP20161,Lemaitre20151,Lemaitre20171,Lemaitre20181,Levashov20111,Levashov2013,Levashov20141,
Levashov20142,Levashov20161,Levashov20162,Levashov20172,Voigtmann2016,Matubayasi2018}.

The considerations of the stress fields from the microscopic perspective often 
utilize essentially the concept of the atomic level stress \cite{Kirkwood19501,Egami19801,Egami19802,Egami19821,Aur19871,Chen19881,Hoheisel19881,Woodcock19911,Woodcock20061,
Goldhirsch2002,Kust2003a,Kust2003b,Picard20041,Leonforte20041,Tanguy20061,Tsamados20081,Leporini20121,HarrowellP20121,ChenYQ2013,
Ashwin20151,IlgP20151,HarrowellP20161,Lemaitre20151,Lemaitre20171,Lemaitre20181,Levashov20111,Levashov2013,Levashov20141,
Levashov20142,Levashov20161,Levashov20162,Levashov20172,Voigtmann2016,Matubayasi2018}. 
This approach, besides addressing the stress states of individual particles, 
also allows addressing the long-range stress correlations.
Considerations of the correlations between the microscopic stress fields
are important because they lead to a better understanding 
of the elastic properties of amorphous materials and also because 
these correlations are related to viscosity via the Green-Kubo expression \cite{HansenJP20061,Green19541,Kubo19571,Helfand19601}.
   
Another widely used approach to address structural/geometrical correlations in 
liquids and glasses is based on the considerations 
of the bond-orientational order (BOO) parameters \cite{Steinhardt19811,Steinhardt19831,Strandburg1992,Tanaka20121}.
In the introduction of this approach \cite{Steinhardt19811,Steinhardt19831}, as in many of its later implementations, 
it was assumed that there is ``a bond" between a pair of particles if these two particles can be considered as the nearest neighbors. 
In the original papers \cite{Steinhardt19811,Steinhardt19831}, the BOO parameters, have been associated with three different groups of bonds. 
In other words, it is possible to say that three different
cases have been considered. In one case, the BOO parameters have been associated with individual
bonds. In another case, the BOO parameters have been associated with a chosen particle and all its nearest
neighbors. In the third case, the BOO parameters have been associated with all bonds in the sample
in order to address the global ``magnetization" of the BOO.
 
In our view, most often the BOO approach is used to demonstrate the development of some local BOO 
in liquids on supercooling through considerations of particles and their nearest-neighbor environments \cite{Steinhardt19811,Steinhardt19831,Strandburg1992,Tomida19951,Dellago20081,Tanaka20121,Patashinki1997,Hirata20131,
Kelton20161,Ryltsev20131,Mecke2013,Mecke20132,HCAndersen1988,Grest1991,Ganesh20051,Zhigilei19941,
Tanaka20101,Stillinger20101,EMa2011,MalinsA20131,Ryltsev20131,Ryltsev20161,Kelton20161}. 
For this, the BOO parameters associated with the spherical harmonics of the order $l=4,6,8,10$
are routinely considered. These choices of $l$ are made because they allow distinguishing between 
the nearest-neighbor cluster geometries in the SC, FCC, HCP, and BCC crystal lattices and also to distinguish 
the icosahedral clusters from the clusters associated with the mentioned lattices.

On the other hand, the nearest-neighbor BOO parameters associated with $l=2$ spherical harmonics 
are closely related to the atomic level stresses, as evident from the considerations 
in Refs. \cite{Steinhardt19811,Steinhardt19831,Chen19881,Tomida19951}.
We will also discuss this connection in this paper.

It is somewhat surprising, but while in the literature there are many discussions of the local and medium range BOO \cite{Steinhardt19811,Steinhardt19831,Strandburg1992,Tomida19951,Dellago20081,Tanaka20121,Patashinki1997,Hirata20131,
Kelton20161,Ryltsev20131,Mecke2013,Mecke20132,HCAndersen1988,Grest1991,Ganesh20051,Zhigilei19941,
Tanaka20101,Stillinger20101,EMa2011,MalinsA20131,Ryltsev20131,Ryltsev20161,Kelton20161}
there appear to be only several publications \cite{Steinhardt19811,Steinhardt19831,Grest1991,Tomida19951,Chen19881} 
in which the long-range correlation functions (CFs), defined through the BOO parameters, have been explicitly considered as a function of distance. 
The reason for this situation might be related to the expectation that for 
the developing icosahedral ordering in liquids on cooling the distance between the two icosahedral clusters 
may not be the best parameter to describe the developing BOO due to the (possibly) branched (fractal) geometry 
of the domains formed by the connected (or interpenetrating) icosahedra.

The basic idea behind this paper is the observation that for the particles
interacting through pair potentials the elementary units of stress are not the particles'  stresses, 
but the stresses associated with interactions between the pairs of particles. 
We gained the idea to consider bonds' stresses from  Ref. \cite{Yoshino20131,Lemaitre20161}. 
There the notion of the bond's stress has been mentioned, though correlations between the bonds' stresses have not been considered. 
Another motivation to consider correlations between the bonds' stresses is based on Ref. \cite{Iwashita20131} 
in which a simple relation between the average lifetime of the nearest-neighbor local atomic environment
and the Maxwell relaxation time has been suggested.  

The primary purpose of this paper, as of our several previous 
publications \cite{Levashov20111,Levashov2013,Levashov20141,Levashov20142,Levashov20161,Levashov20162,Levashov20172}, 
is to develop a better understanding of the atomic scale stress correlations 
relevant to the Green-Kubo expression for viscosity.

The paper is organized as follows. 
In Sec. (\ref{sec:GKexpression}) we discuss the Green-Kubo expression for viscosity from the perspectives
of the atomic level stresses and the bonds' stresses. There we also discuss the rotationally invariant form of the Green-Kubo stress
CF. In Sec. (\ref{sec:BOOA}) we discuss the connection of the microscopic Green-Kubo stress CF in the rotationally invariant form with
the CFs from the BOO approach. The model that has been used in our MD simulations is described 
in Sec. (\ref{sec:model}). In four subsections of Sec. (\ref{sec:results}) we discuss our results.
We conclude in Sec. (\ref{sec:conclusion}).

\section{The Green-Kubo expression for viscosity and microscopic stress tensors \label{sec:GKexpression}}

The Green-Kubo expression is widely used in MD simulations for the calculations of zero-frequency and zero-wavevector viscosity.
It relates shear viscosity to the decay of the macroscopic stress correlation 
function \cite{HansenJP20061,Green19541,Kubo19571,Helfand19601,Hoheisel19881}: 
\begin{eqnarray}
\eta=\frac{V}{k_B T}\int_{0}^{\infty}\left<\Pi^{\alpha \beta}(t_o)\Pi^{\alpha \beta}(t_o+t)\right>_{t_o}dt,
\label{eq:green-kubo-01}
\end{eqnarray}
where $V$ is the volume of the system, $k_B$ is the Boltzmann constant, $T$ is 
the temperature, and $\Pi^{\alpha \beta}(t)$ is the off-diagonal $(\alpha \neq \beta)$ component 
of the macroscopic stress tensor of the system at time $t$. 
The averaging is done over the equilibrium canonical ensemble 
(in practice, over the initial times, $t_o$, under the assumption that ergodicity holds).

The full expression for the macroscopic stress tensor can be found in 
Ref.\cite{HansenJP20061,Green19541,Kubo19571,Helfand19601,Hoheisel19881}. 
We are interested in its simpler form which is the most relevant for the
studies of viscosities of dense and supercooled liquids; i.e.,
we neglect the contributions to the stress tensor associated 
with the velocities of the particles, as it has been shown in multiple previous investigations 
that these contributions are negligibly small ($\approx 2\%$ \cite{Hoheisel19881})
in comparison to the terms involving only interactions between the particles. 
Then the expression for the components of the macroscopic stress tensor can be written as:
\cite{HarrowellP20121,IlgP20151,HarrowellP20161,McQuarrie19761,HansenJP20061,Green19541,
Kubo19571,Helfand19601,Hoheisel19881,Woodcock19911,Woodcock20061,Levashov20111,Levashov2013,Voigtmann2016,Matubayasi2018}:
\begin{eqnarray}
\Pi^{\alpha\beta}(t)= \frac{\rho_o}{N}\sum_{i=1}^{i=N}s_i^{\alpha\beta}(t),\;\;\;\;s_i^{\alpha\beta}\equiv\sum_{j \neq i} 
\varphi'(r_{ij})\frac{r_{ij}^a r_{ij}^b}{r_{ij}},\;\;\;
\label{eq:green-kubo-02}
\end{eqnarray}
where $N$ is the number of particles in the system, $\rho_o \equiv N/V$ is the particles' 
number density, and $\varphi'(r_{ij})$ is the derivative of the pair-interaction 
potential between particles $i$ and $j$.
In the following, we will refer to $s_{i}^{\alpha\beta}$ 
as to (the component of) the atomic stress element of the particle $i$.

Substitution of (\ref{eq:green-kubo-02}) into (\ref{eq:green-kubo-01}) leads to the
following expressions for the CF between the macroscopic stress tensors:
\begin{eqnarray}
&&F^{\alpha\beta}(t)\equiv\left<\Pi^{\alpha \beta}(t_o)\Pi^{\alpha \beta}(t_o+t)\right>_{t_o},
\label{eq:cfa01}\\
&&F^{\alpha\beta}(t)= F_{auto}^{\alpha\beta}(t) + F_{cross}^{\alpha\beta}(t),\label{eq:F-tot-01}\\
&&F_{auto}^{\alpha\beta}(t) \equiv \left<\frac{1}{N}
\sum_{i}s^{\alpha\beta}_i(t_o)s^{\alpha\beta}_i(t_o+t)\right>_{t_o},\label{eq:F-auto-01}\\
&&F_{cross}^{\alpha\beta}(t) \equiv \int_{+\epsilon}^{\infty} \mathcal{F}_{cross}^{\alpha\beta}(t,r)dr,\label{eq:F-cross-01}\\
&&\mathcal{F}_{cross}^{\alpha\beta}(t,r) \nonumber\\
&&\equiv\left<\frac{1}{N}\sum_{i}s^{\alpha\beta}_i(t_o)
\sum_{j\neq i}s^{\alpha\beta}_j(t_o+t)\delta\left(r-r_{ij}(t_o)\right)\right>_{t_o},\label{eq:F-cal-01}
\end{eqnarray}
where $\delta\left(r-r_{ij}(t_o)\right)$ is the $\delta$ function that 
introduces the dependence of the CF (\ref{eq:F-cross-01},\ref{eq:F-cal-01}) on the 
separation, $\vec{r}_{ij}=\vec{r}_j - \vec{r}_i$, between particles $i$ and $j$ at time $t_o$.
In (\ref{eq:F-cross-01}) the lower integral limit is $+\epsilon$ because $r=0$ corresponds to the case
when $i=j$ and this situation is taken into account by $F_{auto}^{\alpha\beta}(t)$ from (\ref{eq:F-auto-01}).

In our previous papers we studied the dependencies of (\ref{eq:cfa01},\ref{eq:F-tot-01},\ref{eq:F-auto-01},\ref{eq:F-cross-01},\ref{eq:F-cal-01}) on $t$ and $r$ \cite{Levashov20111,Levashov2013,Levashov20141,Levashov20142,Levashov20161,Levashov20162,Levashov20172}.
The major result of those investigations, 
in our view, is the demonstration that $F_{auto}(t)$ effectively accounts for
the contribution to viscosity due to the structural relaxation, 
while the contribution to viscosity due to $F_{cross}(t,r)$ is associated with vibrational modes
in liquids and their attenuation. 
Our results suggest that approximately half of the value of viscosity is associated with non-local vibrational modes. 
There are also other papers in which the decomposition of the macroscopic stress correlations 
into the atomic scale stress CFs has been investigated 
\cite{Woodcock19911,Woodcock20061,Lemaitre20151,HarrowellP20161,Voigtmann2016,Lemaitre20171,Lemaitre20181,Matubayasi2018}.
A somewhat different approach, which also addresses the atomic-scale correlations, is based on the introduction 
of a continuous stress field through a coarse-graining procedure \cite{Goldhirsch2002,Lemaitre20151,Lemaitre20171,Lemaitre20181}. 
Without going into the details, it is possible to say that the existence of nonlocal stress fields has been demonstrated 
and their structure and time evolution has been addressed.

The major idea behind this paper is quite simple. 
It is easy to notice that (\ref{eq:green-kubo-02}) can be rewritten 
as follows:
\begin{eqnarray}
&&\sigma^{\alpha\beta}(t)= -\frac{2\rho_o}{N}\sum_{ij} f_{ij}^{\alpha}(t) r_{ij}^{\beta}(t)=-\frac{2\rho_o}{N}\sum_{j \neq i} b_{ij}^{\alpha \beta}(t),\;\;\label{eq:green-kubo-02x}\\
&&b_{ij}^{\alpha\beta}(t) \equiv f_{ij}^{\alpha}(t) r_{ij}^{\beta}(t),\;\;\;\;b_{ij} \equiv f_{ij}r_{ij} \equiv |\vec{f}_{ij}|\cdot |\vec{r}_{ij}|,
\label{eq:green-kubo-02y}
\end{eqnarray}
where $f_{ij}^{\alpha} = - (d\phi_{ij}/dr_{ij})(r_{ij}^{\alpha}/r_{ij})$ is the $\alpha$ component of the force acting 
on particle $i$ from particle $j$ and $b_{ij}^{\alpha \beta}$ is the $\alpha\beta$ component of the stress tensor associated with 
the interaction between particles $i$ and $j$.
The factor of $2$ in (\ref{eq:green-kubo-02x}) originates from the fact that every pair of particles in (\ref{eq:green-kubo-02}) is counted twice,
while in (\ref{eq:green-kubo-02x}) every pair of particles is counted only once.
Note that if particles $i$ and $j$ do not interact, i.e., they are too far away from each other,
then $f_{ij}^{\alpha}=0$ and $b_{ij}^{\alpha \beta}=0$.

If the interaction between the particles is such that the first nearest neighbors are well defined and 
the interaction with the second neighbors is negligibly small in comparison to the interaction with 
the first neighbors then it is possible to think about the $b_{ij}^{\alpha \beta}$ for the nearest 
neighbors $i$ and $j$ as about the stress tensor of the $ij$-bond. 
We will assume, as is done usually, that the $ij$ bond is located at 
${\vec{R}}_{ij} \equiv ({\vec{r}}_{i}+{\vec{r}}_{j})/2$. 

In the following discussions, we use the 
terms ``bond's stress" and ``bond-stress correlation function" to describe
structural correlations in a model liquid at the atomic scale.
The use of this terminology does not imply that we assign a precise physical meaning to the
concept of the ``bond's stress" in the way it is done for the macroscopic 
stress tensor in the continuum theory of elasticity. 
From the perspective of addressing the microscopic structure of liquids, 
our use of these terms is essentially identical to the widely used terminology associated
with the concept of atomic level stresses \cite{ChenYQ2013,HarrowellP20161,Woodcock19911,Woodcock20061,Tsamados20081,Egami19801,Egami19802,Egami19821,Chen19881,
Kust2003a,Kust2003b,Levashov20111,Levashov20141,Levashov20161,Levashov20162,Levashov20172,Voigtmann2016,Matubayasi2018}. 

Using the components of the pair interaction stress tensors, $b_{ij}^{\alpha \beta}$, we can rewrite 
the CF of the macroscopic stress tensors in (\ref{eq:green-kubo-01},\ref{eq:cfa01}) similarly to expressions
(\ref{eq:cfa01},\ref{eq:F-tot-01},\ref{eq:F-auto-01},\ref{eq:F-cross-01},\ref{eq:F-cal-01}):
\begin{eqnarray}
&&B^{\alpha\beta}(t) \equiv F^{\alpha\beta}(t) \equiv \left<\Pi^{\alpha \beta}(t_o)\Pi^{\alpha \beta}(t_o+t)\right>_{t_o},\label{eq:cfb01}\\
&&B^{\alpha\beta}(t)=B_{auto}^{\alpha\beta}(t) + B_{cross}^{\alpha\beta}(t),\label{eq:B-tot-01}\\
&&B_{auto}^{\alpha\beta}(t) \equiv \left<\frac{1}{N}\sum_{ij}b^{\alpha\beta}_{ij}(t_o)b^{\alpha\beta}_{ij}(t_o+t)\right>_{t_o},\label{eq:B-auto-01}\\
&&B_{cross}^{\alpha\beta}(t) \equiv \int_{0}^{\infty} \mathcal{B}_{cross}^{\alpha\beta}(t,r)dr,\label{eq:B-cross-01}\\
&&\mathcal{B}_{cross}^{\alpha\beta}(t,r)\equiv\nonumber\\
&&\left<\frac{1}{N}\sum_{ij}b^{\alpha\beta}_{ij}(t_o)\sum_{kh}b^{\alpha\beta}_{kh}(t_o+t)\delta\left(r-r_{ij,kh}\right)\right>_{t_o},\label{eq:B-cal-01}\\
&&r_{ij,kh}\equiv |{\vec{R}}_{kh}(t_o+t)-{\vec{R}}_{ij}(t_o)|.
\end{eqnarray}
Note that the notation $r_{ij,kh}$ is for the distance between 
the bond $ij$ at time $t_o$ and the bond $kh$ at time $t_o+t$.
Note also that in order to obtain the correct value of the correlation 
function $B_{cross}^{\alpha\beta}(t)$ in (\ref{eq:B-cross-01}), 
we simply need to count correlations between all interacting pairs, 
i.e., the definition of ${\vec{R}}_{ij}$ does not affect the value of 
the macroscopic CF. 

The number of interacting pairs in the system fluctuates.
For this reason the normalization in Eqs. (\ref{eq:B-auto-01}) and (\ref{eq:B-cross-01}) is to the number
of particles in the system.
Of course, definitions (\ref{eq:F-auto-01}),(\ref{eq:F-cross-01}) 
and
(\ref{eq:B-auto-01}),(\ref{eq:B-cross-01}) 
should lead to exactly the same result, i.e.,
to the value of the macroscopic stress tensor.

In the preceding paragraph we made a reference to the exactly the same values of the CF between 
the macroscopic stress tensors which should be obtained from the microscopic approaches based on considerations
of the atomic stresses or bond stresses. In our view, it is worth making here a comment 
related to the history of application of the Green-Kubo expression. 
As far as we understand, the original derivations of the Green-Kubo 
expression \cite{Green19541,Kubo19571,Helfand19601}
are actually microscopic and thus microscopic considerations adopted 
relatively recently \cite{Woodcock19911,Woodcock20061,Levashov20111,Levashov2013,Levashov20141,Levashov20142,
Levashov20161,Levashov20162,Levashov20172,Lemaitre20151,HarrowellP20161,Voigtmann2016,Lemaitre20171,Lemaitre20181,Matubayasi2018}
are much closer in spirit to the derivations of the Green-Kubo expression  
than the macroscopic view of the Green-Kubo expression usually used. 
Our research of the literature suggests that the macroscopic view of 
the microscopically-derived Green-Kubo expression has been
adopted in one of the first papers on viscosity calculations in computer simulations \cite{Verlet19731} (see also Ref. \cite{Alder1970}). 
There the connection between the microscopic and macroscopic views has not been addressed in details. 
The issue concerning the possible non-equivalence of the microscopic and macroscopic perspectives has been
discussed at first in Ref. \cite{Erpenbeck19951}. 
Thus, in our view, it is important to realize that microscopic considerations are more original
than the macroscopic considerations. 
The exact equivalence between the macroscopic and microscopic perspectives in the case of computer simulations
on finite systems with the periodic boundary conditions follows exactly from the fact that the considered systems are finite
and that there are the periodic boundary conditions. 
 
Any physically meaningful structural CFs describing isotropic states
should not depend on the orientation of the observation coordinate frame.
This means that (time or Gibbs-ensemble averaged) CFs (\ref{eq:F-auto-01},\ref{eq:F-cross-01},\ref{eq:F-cal-01}) and 
(\ref{eq:B-auto-01},\ref{eq:B-cross-01},\ref{eq:B-cal-01}) should be the same in any observation coordinate frame. 
Therefore, we may think that the averaging over $t_o$ also includes
in itself the averaging over all possible orientations of the observation coordinate frame.
Thus, instead of considering the contribution of every product $s^{\alpha\beta}_i s^{\alpha\beta}_j$ to 
(\ref{eq:F-auto-01},\ref{eq:F-cross-01},\ref{eq:F-cal-01}) or $b^{\alpha\beta}_{ij} b^{\alpha\beta}_{kh}$ 
to (\ref{eq:B-auto-01},\ref{eq:B-cross-01},\ref{eq:B-cal-01}) in a particular observation coordinate frame, 
we can instead consider their contributions averaged 
over all directions of the observation coordinate frame. 
Note in this context that the product $s^{\alpha\beta}_i s^{\alpha\beta}_j$ 
is not rotationally invariant. 
More detailed considerations of the relevant issues have 
been presented in Ref. \cite{Levashov20161,Levashov20162,Lemaitre20151,Lemaitre20171,Lemaitre20181}
Earlier the relevant issues actually have been  
addressed within the bond-orientational order approach through considerations 
of the rotationally invariant combinations of the  BOO parameters \cite{Steinhardt19811,Steinhardt19831,Strandburg1992,Tanaka20101}.

In Ref.\cite{Levashov20162} we expressed the value 
of $s^{\alpha\beta}_i s^{\alpha\beta}_j$ averaged
over all directions of the observation coordinate frame in terms of the stress 
components in a particular observation coordinate frame. 
In particular, it has been shown that (see expression (56) in Ref. \cite{Levashov20162}):
\begin{eqnarray}
\left\langle s_i^{xy} s_j^{xy}\right\rangle_{\Omega} = &&-\left(\frac{3}{10}\right)p_i p_j +
\left(\frac{1}{10}\right)\sum_{n,m=1}^{n,m=3}\lambda_i^n \lambda_j^m \left(c_{ij}^{nm}\right)^2 \nonumber\\
= && \left(\frac{1}{10}\right)\sum_{n,m=1}^{n,m=3}\lambda_i^n \lambda_j^m \left[\left(c_{ij}^{nm}\right)^2-\frac{1}{3}\right],
\label{eq:Gs}
\end{eqnarray}
where $\left\langle ... \right\rangle_{\Omega}$ is the averaging over the directions of the observation coordinate frame. 
$p_i\equiv \tfrac{1}{3}\left(\lambda_i^1+\lambda_i^2+\lambda_i^3\right)$ is the atomic scale pressure on particle $i$,
$\lambda_i^1$ is the first eigenvalue of the stress tensor $\underline{\underline{s}}_i$, 
e.t.c. $c_{ij}^{nm}$ is the cosine of the angle
between the $n$-th eigenvector of the tensor $\underline{\underline{s}}_i$ and $m$-th eigenvector of the tensor $\underline{\underline{s}}_j$. 

In the considerations of the stress CFs in isotropic liquids one can expect that the right-hand side of (\ref{eq:Gs})
might also depend on the orientations of the eigenvectors of the stress tensors with respect to the direction from
$i$ to $j$. Indeed, for the stress tensors of the particles $i$ and $j$ it is the only special direction in an isotropic liquid.
However, note that it follows from (\ref{eq:Gs}) that only the orientations of the stress tensor eigenvectors with respect 
to each other are relevant, while the orientations of the eigenvectors with respect to the direction 
from $i$ to $j$ are irrelevant. See Ref. \cite{Levashov20161,Levashov20162,Lemaitre20151,Lemaitre20171,Lemaitre20181} and also expression
(12) from Ref.\cite{Chen19881}.

\begin{figure}
\begin{center}
\includegraphics[angle=0,width=2.5in]{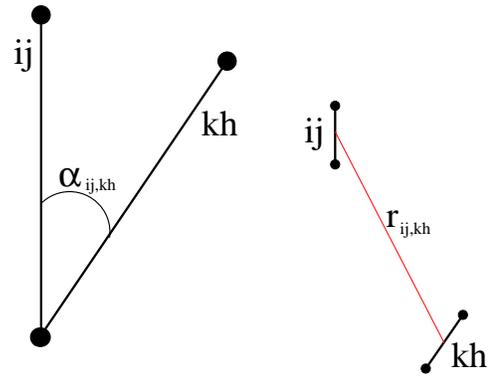}
\caption{Let us suppose that bonds $ij$ and $kh$ are at distance, $r_{ij,kh}$, 
from each other and that the angle between their directions is $\alpha_{ij,kh}$.
If there are no correlations between the directions of the bonds at distance
$r_{ij,kh}$ then the function 
$f(r) \equiv  \left\langle\left[\cos^2[\alpha_{(ij,kh)}] - 1/3\right] \delta(r-|{\bf{r}}_{ij}-{\bf{r}}_{kh}|)\right\rangle_{(ij,kh)}$ 
should be equal to zero for the separation distance $r_{ij,kh}$. 
As discussed in the text, 
the average value of this function is directly related to viscosity.
}\label{fig:bonds-geometry}
\end{center}
\end{figure}

Clearly, (\ref{eq:Gs}) can also be applied to the stress tensors of the bonds. 
The stress tensor of any $ij$ interacting pair of particles has only one nonzero eigenvalue, i.e., $b_{ij}=f_{ij}r_{ij}$. 
Thus, for the bonds expression (\ref{eq:Gs}) is very simple:
\begin{eqnarray}
&\left\langle b_{ij}^{xy} b_{kh}^{xy}\right\rangle_{\Omega} = 
\left(\frac{1}{10}\right)b_{ij}b_{kh}\left[\cos^2\left(\alpha_{ij,kh}\right)-\tfrac{1}{3}\right].\;\;\;
\label{eq:Gb}
\end{eqnarray}
See Fig. \ref{fig:bonds-geometry}.
With (\ref{eq:Gb}) expression (\ref{eq:B-cal-01}) becomes a special pair density function of the bonds, 
\begin{eqnarray}
&&\mathcal{B}(t,r)\equiv \frac{1}{10N}\nonumber\\
&&\cdot\left<\sum_{ij \neq kh}b_{ij}b_{kh}\left[\cos^2\left(\alpha_{ij,kh}\right)-\tfrac{1}{3}\right]\delta\left(r-r_{ij,kh}\right)\right>_{t_o},\;\;\;\;\;\label{eq:B-cal-2}
\end{eqnarray}
in which the contribution of every pair of bonds is weighted by their tensions 
and the mutual orientation factor $\left[\cos^2\left(\alpha_{ij,kh}\right)-\tfrac{1}{3}\right]$.
See Fig. \ref{fig:bonds-geometry}.
It is easy to see that if in 3D there is no correlation in the orientations 
of the bonds at distance $r$ then 
(\ref{eq:Gb}) averages to zero.

\section{On the connection with the Bond-Orientational Order Approach \label{sec:BOOA}}

The changes in the structures of liquids on cooling are often described 
within the bond-orientational order (BOO) approach 
\cite{Steinhardt19811,Steinhardt19831,Strandburg1992,
Tomida19951,Dellago20081,Stillinger20101,Tanaka20121}. 
The stress CFs (\ref{eq:cfa01},\ref{eq:F-tot-01},\ref{eq:F-auto-01},\ref{eq:F-cross-01},\ref{eq:F-cal-01}
\ref{eq:cfb01},\ref{eq:B-tot-01},\ref{eq:B-auto-01},\ref{eq:B-cross-01},\ref{eq:B-cal-01}) that we consider 
are actually closely related to some of the CFs that have been defined within the BOO approach.

The basic quantities introduced in the BOO approach are the spherical harmonics
associated with ``a bond" connecting a pair of particles \cite{Steinhardt19811,Steinhardt19831,Strandburg1992}. 
In most cases, it is assumed that particles $i$ and $j$ are connected by a bond if they are 
within some distance from each other. 
It has been discussed recently that this simple definition has shortcomings and the possible 
fixes have been suggested \cite{Mecke2013,Mecke20132}. 
In our view, the considerations of the bonds' stresses discussed here  
also touch on the issues raised in \cite{Mecke2013,Mecke20132}. 

In any case, for now, we assume that we can associate 
``a bond" with particles $i$ and $j$ if they are within some distance
from each other. The direction of this bond, $\vec{r}_{ij} \equiv \vec{r}_j-\vec{r}_i$, 
can be characterized with ``the bond orientation parameters", $Q_{lm}(\vec{r}_{ij})$,
which are just the spherical harmonics, $Y_{lm}\left(\theta_{ij},\phi_{ij}\right)$:
\begin{eqnarray}
Q_{lm}(\vec{r}_{ij}) \equiv Y_{lm}\left(\theta_{ij},\phi_{ij}\right).
\label{eq:boo-Qlm}
\end{eqnarray}
In the following, as before, we assume that the bond $\vec{r}_{ij}$ 
is located at $\vec{R}_{ij}\equiv \left(\vec{r}_i+\vec{r}_j\right)/2$.
Note that the values of $Q_{lm}(\vec{r}_{ij})$ depend on the choice 
of the observation coordinate frame.

Further, ``the bond orientation parameters" for some groups of bonds are usually introduced:
\begin{eqnarray}
\overline{Q}_{lm}(group) \equiv \left< Q_{lm}(\vec{r}) \right>_{group},
\label{eq:boo-Qlm-group}
\end{eqnarray}
where the angular brackets on the right-hand side signify
the averaging over the selected group of bonds.

Further, in order to avoid the dependence of $\overline{Q}_{lm}$
on the choice of the observation coordinate frame, 
the rotationally invariant combinations of $\overline{Q}_{lm}$
are introduced:
\begin{eqnarray}
Q_{l} \equiv \left[\frac{4\pi}{2l+1} \sum_{m=-l}^{m=+l} \left|\overline{Q}_{lm}\right|^2 \right]^{1/2},
\label{eq:boo-Ql}
\end{eqnarray}
\begin{eqnarray}
W_{l} \equiv  \sum_{\substack{m_1,m_2,m_3 \\ m_1+m_2+m_3=0}} \left(\Gj{l}{l}{l}{m_1}{m_2}{m_3}\right) \overline{Q}_{l m_{1}}\overline{Q}_{l m_{2}}\overline{Q}_{l m_{3}},\;\;\;\;\;\;
\label{eq:boo-Wl}
\end{eqnarray}
where right after the sum sign in (\ref{eq:boo-Wl}) stand(s) Wigner 3j-symbol(s).

Considerations of the ``bond-order parameters", $Q_l$ and $W_l$, for the particles in simple lattices, 
such as simple cubic (SC), face-centered cubic (FCC), or body-centered cubic (BCC) allow one
to distinguish these structures from each other and also, 
in particular, to distinguish particles with the icosahedral environment
from the particles with the environments characteristic for the mentioned crystal 
lattices \cite{Steinhardt19811,Steinhardt19831,Strandburg1992,
Stillinger20101,Tanaka20121}. 
At present, in our view, the BOO approach is used most frequently to characterize the geometry of 
the nearest-neighbor shells around the chosen particles and the formation of domains from the clusters of a particular (usually icosahedral)  symmetry \cite{Steinhardt19811,Steinhardt19831,Strandburg1992,Tomida19951,Dellago20081,Tanaka20121,Patashinki1997,Hirata20131,
Kelton20161,Ryltsev20131,Mecke2013,Mecke20132,HCAndersen1988,Grest1991,Ganesh20051,Zhigilei19941,
Tanaka20101,Stillinger20101,EMa2011,MalinsA20131,Ryltsev20131,Ryltsev20161,Kelton20161}. 

It has been demonstrated that in order to distinguish between different crystalline (and icosahedral)
motives it might be better to consider as groups of bonds all bonds 
associated with a chosen particle and its nearest neighbors,
i.e., also include in the group all bonds associated with the nearest neighbors \cite{Dellago20081}.

The BOO approach can also be used to address correlations between the local environments of different particles 
\cite{Steinhardt19811,Steinhardt19831,Grest1991,Tomida19951,Chen19881}.
For this, in Refs. \cite{Steinhardt19811,Steinhardt19831,Grest1991} the following 
rotation-invariant CF has been considered:   
\begin{eqnarray}
G_l(r)\equiv &&\left(\frac{4\pi}{2l+1}\right)\left(\frac{1}{G_o(r)}\right) \cdot \nonumber\\
&&\cdot\sum_{m=-l}^{m=+l}\left< Q_{lm}(\vec{r}_{ij})Q_{lm}^{*}(\vec{r}_{kh})\delta(r-r_{ij,kh}) \right>_{ij,kh},\;\;\;\;\;\;
\label{eq:Glr-01}
\end{eqnarray}
where the averaging in (\ref{eq:Glr-01}) is effectively over all pairs of 
bonds, i.e., $(ij)$ and $(kh)$, distance $r$ away from each other.
In (\ref{eq:Glr-01}) the function $G_o(r)$ is the pair density function of the bonds, whose 
definition also follows from (\ref{eq:Glr-01}). However, in defining $G_o(r)$ through (\ref{eq:Glr-01})
there should not be $G_o(r)$ in the denominator on the right-hand side.

While the BOO approach can be used to address the long-range correlations between the bonds and the small clusters 
of particles it appears that the number of studies in which this has been done is 
relatively small \cite{Steinhardt19811,Steinhardt19831,Grest1991,Tomida19951,Chen19881}.
In such considerations the spherical harmonics corresponding 
to $l=4$ and $l=6$ are usually considered due to the search for a proliferating icosahedral order.
In Ref. \cite{Chen19881} has been considered a triple-product CF associated with the spherical harmonics 
of the order $l=2$ for the two clusters of particles and with the spherical harmonics of 
the order $l=2$, $l=4$, and $l=6$ for the radius vector between the two clusters.
However, we have not found studies in which the considerations of the orientations 
of the individual bonds described through the binary products of the $l=2$ spherical harmonics have been presented. 

Let us consider two bonds, $(ij)$ and $(kh)$, with the angle $\alpha_{ij,kh}$ between them. 
See Fig. \ref{fig:bonds-geometry}.
For these two bonds, the addition theorem for the spherical harmonics reads as follows:
\begin{eqnarray}
&&P_l(\cos\left(\alpha_{ij,kh}\right))\nonumber\\
&&=\frac{4\pi}{2l+1}\sum_{m=-l}^{m=+l}Q_{lm}(\theta_{ij},\phi_{ij})Q_{lm}^{*}(\theta_{kh},\phi_{kh}),
\label{eq:spherical-01}
\end{eqnarray}
where $P_l(\cos\left(\alpha_{ij,kh}\right))$ is the Legendere polynomial of degree $l$.
Thus, (\ref{eq:Glr-01}) can be rewritten as (it is clear that the summation over $m$ 
and the averaging over different bonds can be changed in order): 
\begin{eqnarray}
G_l(r)\equiv \tfrac{1}{G_o(r)}
\left<P_l(\cos\left(\alpha_{ij,kh}\right))\delta(r-r_{ij,kh})\right>_{ij,kh},\;\;\;\;\;\;
\label{eq:Glr-02}
\end{eqnarray}
In particular, for $l=2$,
\begin{eqnarray}
P_2(\cos\left(\alpha_{ij,kh}\right))=\tfrac{3}{2}\left[\cos^2\left(\alpha_{ij,kh}\right)-\tfrac{1}{3}\right].
\label{eq:legendre2}
\end{eqnarray}
and (\ref{eq:Glr-01}) can be rewritten as
\begin{eqnarray}
G_2(r)\equiv \tfrac{3}{2G_o(r)}\cdot\left<\left[\cos^2\left(\alpha_{ij,kh}\right)-\tfrac{1}{3}\right]\delta(r-r_{ij,kh})\right>_{ij,kh}.\;\;\;\;\;\;
\label{eq:Glr-03}
\end{eqnarray}

The comparison of (\ref{eq:Glr-03}) with (\ref{eq:B-cal-2}) shows that the bond-stress CF (BSCF) associated 
with viscosity, (\ref{eq:B-cal-2}), is closely related to the $l=2$ bond-order CF (\ref{eq:Glr-03}). 
The differences between the two CFs are associated with the tensions 
of the bonds in (\ref{eq:B-cal-2})
and with the normalization of the bond-order CF to $(\sim 1/G_o(r))$.

Considerations of the bond-order correlations associated with all nearest neighbors of particles $i$ and $k$
can also be easily done with expression (\ref{eq:Glr-02}). For this it is necessary to introduce into 
(\ref{eq:Glr-02}) the summations over the particles $j$ and $h$ and write in the $\delta$ function 
$r_{ij}$ instead of $r_{ij,kh}$.

Rewriting expression (\ref{eq:Glr-01}) in the form of (\ref{eq:Glr-02}) is, of course, a trivial point.
However, we find it somewhat puzzling that the form (\ref{eq:Glr-02}) usually is not considered in the literature.
In our view, expression (\ref{eq:Glr-02}) provides a simple and intuitive insight into the geometrical nature of the correlations
behind the expression (\ref{eq:Glr-01}). In particular, expression (\ref{eq:Glr-02}) explicitly shows how 
the functions $G_l(r)$ for $l>2$ depend on the angles between the bonds, 
though in those expressions the angles between the bonds enter through more complex higher degree
Legendre polynomials. Note that the direction from one bond to another is irrelevant for all $l$. 
From this perspective, it is also of interest to gain some intuitive insight into the nature of geometrical
correlations behind expression (\ref{eq:boo-Wl}). An illustrative example on this issue is given in the Appendix.

\section{The model \label{sec:model}}

In order to address the behavior of correlation
function $\mathcal{B}_{cross}^{\alpha\beta}(t,r)$ (\ref{eq:B-cross-01})
we used a binary equimolar system of particles interacting through purely repulsive potential(s):
\begin{eqnarray}
\phi_{ab}(r)=\epsilon\left(\frac{\sigma_{ab}}{r}\right)^{12},\;\;\;
\label{eq:potential}
\end{eqnarray}
where ``$a$" and ``$b$" mark the types of particles: ``$A$" or ``$B$". 
The values of the parameters are $\sigma_{AA}=1.0$, $\sigma_{AB}=1.1$, $\sigma_{BB}=1.2$.
The masses of the particles are $m_{A}=1.0$, $m_{B}=2.0$. 
The chosen value of the particles number density is $\rho_o = (N_{A}+N_{B})/V =0.80$.

In the following, the distance is measured in the units of $\sigma \equiv \sigma_{AA}$ and 
temperature in the units of $\epsilon$. 
The unit of time is $\tau = (m\sigma^2/\epsilon)^{1/2}$.

In our simulations, in order to address possible size effects, we considered the systems of two sizes. 
The small and the large systems contained 5324 and 62500 particles in total correspondingly.
The half lengths of the edges of the cubic simulation boxes 
were $(L/2) \cong 9.41 \sigma$ and $(L/2) \cong 21.38 \sigma$ correspondingly.
The periodic boundary conditions in $\hat{x},\hat{y},\hat{z}$ directions have been applied.

Previously we already used this model to address the behavior of 
the atomic-stress correlation function (ASCFs) $F_{auto}^{\alpha\beta}(t)$ and $\mathcal{F}_{cross}^{\alpha\beta}(t,r)$ 
(\ref{eq:F-auto-01},\ref{eq:F-cross-01}) \cite{Levashov20162,Levashov20172}.
This and similar models also have been used by other authors to address
certain features of supercooled liquids \cite{Hansen19881,Mizuno20111,Mizuno20131}.

The simulations have been performed using the LAMMPS molecular dynamics package \cite{Plimpton1995,lammps}.

The methodological details concerning the system preparation and equilibration 
can be found in Ref. \cite{Levashov20162}.
In order to collect sufficient statistics for the BSCF for $t=0$ at $T=0.26$ 
on the system with the total number of particles $N= 62500$
we considered the structures from 3 consecutive MD runs.
In every run we generated 100 structures. 
The separation time between the two consecutive saved structures was $10\tau$. 
See Fig.2 of Ref.\cite{Levashov20162}.  
We found only small statistical differences between the results from these 3 MD runs.
The curves presented in this paper were produced by averaging the data from these 3 MD runs.
The statistics for the other temperatures and times were obtained on approximately the same amount of data.

\section{Results \label{sec:results}}

\subsection{The average correlation functions}

In this paper, we consider the model liquid at two temperatures, i.e., at $T=1.00$ and $T=0.26$.
The temperature $T=1.00$ approximately corresponds to the potential energy landscape crossover temperature, 
while at $T=0.26$ the liquid is in the deeply supercooled state.
See Ref. \cite{Levashov20162,Levashov20172} for the relevant temperature scales and the results for the ASCFs: 
$F(t)$, $F_{auto}(t)$, $F_{cross}(t)$, $\mathcal{F}_{cross}(t,r)$ 
(\ref{eq:cfa01},\ref{eq:F-tot-01},\ref{eq:F-auto-01},\ref{eq:F-cross-01},\ref{eq:F-cal-01}).

\begin{figure}
\begin{center}
\includegraphics[angle=0,width=3.3in]{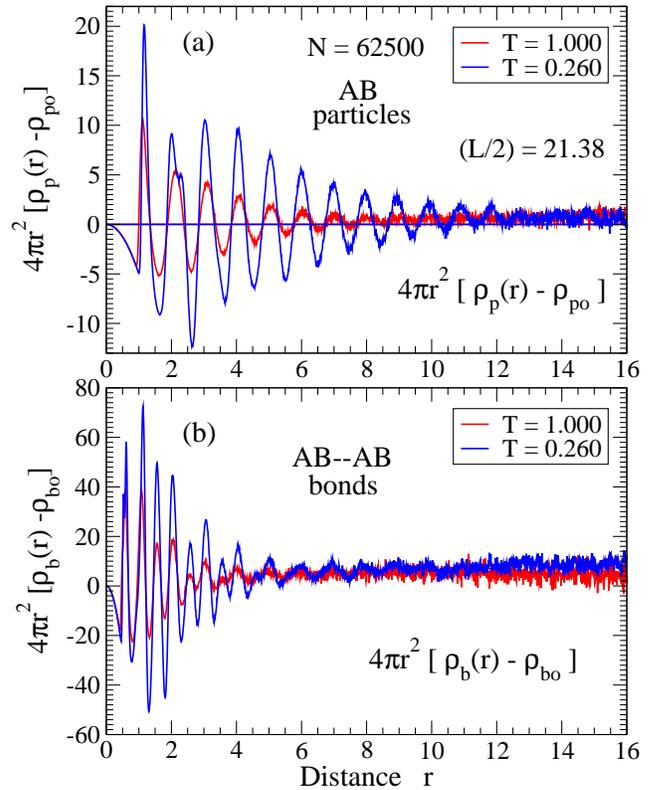}
\caption{The top panel shows the correlation 
function $4\pi r^2 [\rho_p(r)-\rho_{po}]$ between the particles of different types.
Note the presence of the distance-dependent scaling factor $4\pi r^2$. 
The data were obtained on the system with $(L/2) \approx 21.38$.
The bottom panel shows the CF $4\pi r^2 [\rho_b(r)-\rho_{bo}]$ between the
``AB"-``AB" bonds. The notation $\rho_{bo}$ is used for the average number
density of such bonds in the system.
}\label{fig:pdfpb}
\end{center}
\end{figure}

In Fig.\ref{fig:pdfpb} we show the results for the $4 \pi r^2$ scaled 
pair density functions for the particles and for the bonds; i.e.,
we consider the functions
\begin{eqnarray}
&&r \cdot G_p(r) \equiv 4\pi r^2 \left[\rho_p(r) - \rho_{po}\right], \label{eq:pdfparti}\\
&&r \cdot G_b(r) \equiv 4\pi r^2 \left[\rho_b(r) - \rho_{bo}\right], \label{eq:pdfbonds}
\end{eqnarray}
where $G_p(r)$ and $G_b(r)$ are the pair distribution functions (PDFs) for the particles and the bonds.

In the literature the functions $\rho_p(r)$ or $G_p(r)$ are usually considered since 
$\rho_p(r)$ has the clear physical meaning while $G_p(r)$ can be obtained by Fourier
transform from the experimental scattering intensity \cite{HansenJP20061}. 
We, however, consider here the CFs with the additional $r$ factor because
these functions allow making more direct comparisons with the stress CFs
(\ref{eq:cfa01},\ref{eq:F-tot-01},\ref{eq:F-auto-01},\ref{eq:F-cross-01},\ref{eq:cfb01},\ref{eq:B-tot-01},\ref{eq:B-auto-01},\ref{eq:B-cross-01}) 
which are directly relevant to viscosity. 
Note again that CFs (\ref{eq:F-cal-01},\ref{eq:B-cal-01}) contain
in themselves the correlation of a chosen particle (bond) with all other particles (bonds) 
at some distance from it, i.e., they also include in themselves the factor $4\pi r^2$.

It follows from both panels of Fig.\,\ref{fig:pdfpb} that one can clearly distinguish 10 (or 12) coordination 
shells in the $4\pi r^2$ scaled pair density functions for the particles and for the bonds at $T=0.260$. 
At the high temperature, $T= 1.0$, one can distinguish $\sim 8$ coordination shells for the particles and $\sim 7$ 
coordination shells for the bonds. Of course, it would be impossible to observe that many coordination shells without the $4\pi r^2$ scaling. 
Note that the data have been obtained on the cubic system
with $(L/2)=21.38$ with the periodic boundary conditions in $\hat{x},\hat{y},\hat{z}$ directions; 
i.e., note that the considered CFs almost completely decay on the length scales smaller than $(L/2)$.

\begin{figure}
\begin{center}
\includegraphics[angle=0,width=3.3in]{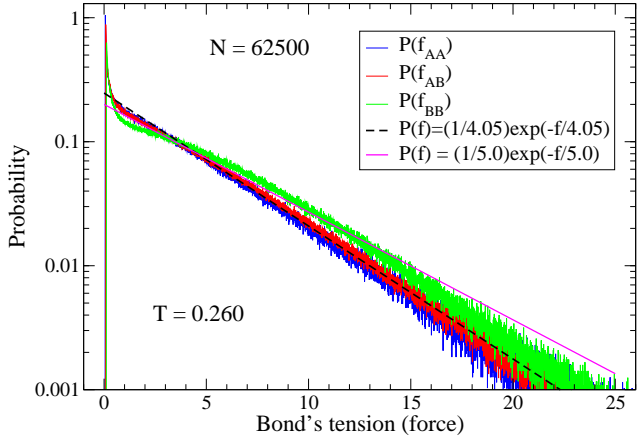}
\caption{The probability distributions of the force tensions between "AA", "AB", and "BB"
particles at $T=0.260$ for the system of 62500 particles. Note that for the particles
separated by the distance corresponding to the position of the first peak in the PDF the repulsive
force is $\approx 12$.
}\label{fig:force-distribution}
\end{center}
\end{figure}

Figure \ref{fig:force-distribution} shows the probability distributions of the bonds' tensions between ``AA", ``AB", and ``BB" particles. 
As follows from the figure, these distributions are close to exponentials, 
as expected according to Ref. \cite{OHern20011,OHern20021,Glotzer20041,Braka20111,Boberski20131}.

\begin{figure}
\begin{center}
\includegraphics[angle=0,width=3.3in]{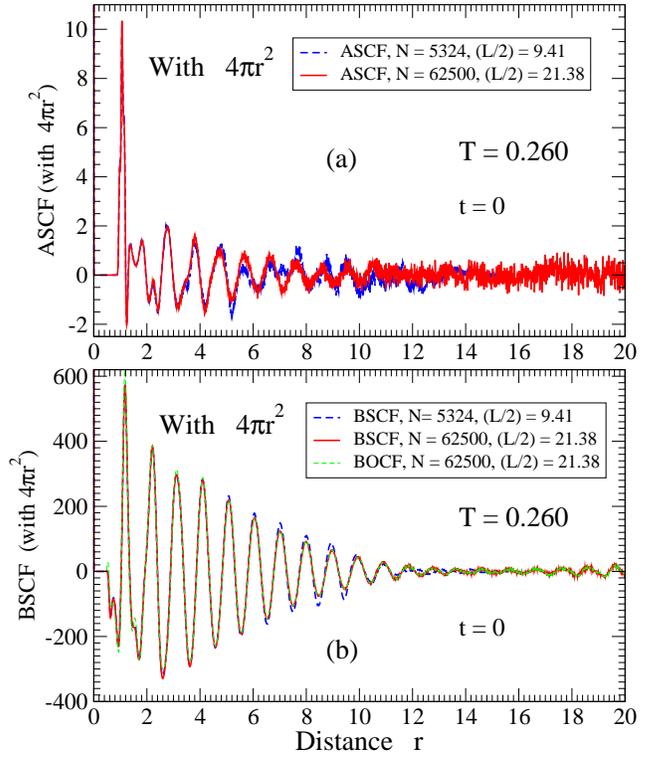}
\caption{Panel (a) shows the ASCF $\mathcal{F}_{cross}^{xy}(r)$ (\ref{eq:F-cal-01}) 
between the particles of types ``A" and ``B" at time $t=0$ and $T=0.260$ for the systems of two sizes. 
Panel (b) shows the BSCF $\mathcal{B}_{cross}^{xy}(r)$ (\ref{eq:B-cal-01}) 
at $t=0$ and $T=0.260$ between the bonds of all types.
The results from the systems of two sizes are shown.
It also shows the bond-order CF scaled by (29.22).
The long-range oscillations in the curves suggest the existence of some long-range BOO.
The similarity between the BSCF and the bond-order CF show that the oscillations 
in the BSCF primary reflect the existence of some BOO.  
Note that at large distances oscillations in the BSCF are more 
pronounced and more regular than the oscillations in the ASCF.
}\label{fig:sscf-at-bs}
\end{center}
\end{figure}

Figure \ref{fig:sscf-at-bs}(a) shows ``the total" ASCFs
(\ref{eq:F-cross-01}) for the systems with $(L/2) = 9.41$ and $(L/2) = 21.38$ at time $t=0$. 
``The total" means that the shown CFs are the sums of the CFs
between ``AA", ``AB", and ``BB" particles. 
The behavior of such and closely related CFs has been studied in 
Ref.\cite{Woodcock19911,Woodcock20061,Levashov20111,Levashov2013,Levashov20141,Levashov20142,
Levashov20161,Levashov20162,Levashov20172,Lemaitre20151,HarrowellP20161,Voigtmann2016,Lemaitre20171,Lemaitre20181,Matubayasi2018}.

Figure \ref{fig:sscf-at-bs}(b) shows the BSCFs calculated from the same 
structural data that have been used to produce Fig.\,\ref{fig:sscf-at-bs}(a). 
The contributions from all bonds, i.e., ``AA", ``AB", and ``BB" 
have been taken into account. Note rather different scales
on the $y$ axes in Fig.\,\ref{fig:sscf-at-bs}(a) and Fig.\,\ref{fig:sscf-at-bs}(b)

It is clear from the comparison of Fig.\,\ref{fig:sscf-at-bs}(a) with Fig.\,\ref{fig:sscf-at-bs}(b) 
that the long-range structural correlations are more pronounced and more regular in 
Fig.\,\ref{fig:sscf-at-bs}(b), i.e., in the BSCFs.
On the other hand, it is possible to think that the ASCFs 
in Fig. \ref{fig:sscf-at-bs}(a) contain some fine features--such as the splitting of the second peak--which are not
present or not well pronounced in the BSCFs.

In Fig.\,\ref{fig:sscf-at-bs}(b) the bond-order CF is also shown. 
It has been calculated similarly to the BSCF, i.e., according 
to (\ref{eq:B-cal-2}), 
but in the calculations of the bond-order CF it has been assumed 
that $b_{ij}=f_{ij}r_{ij}=1$ for all interacting pairs 
that have been counted as the bonds. 
The cutoff distances for the bonds' assignments 
for all pair types have been chosen to correspond to the first
minimums in the corresponding partial PDFs. 
The bond-order CF obtained in this way has been multiplied by a constant 
scaling factor of $29.22$ in order to make a comparison with the BSCF. 
It follows from the figure that the scaled bond-order CF and the BSCF almost coincide. 
This shows that the BSCF describes mostly the BOO, while differences 
in the tensions of the bonds are not that important
for the structure of the BSCF.  

\begin{figure}
\begin{center}
\includegraphics[angle=0,width=3.3in]{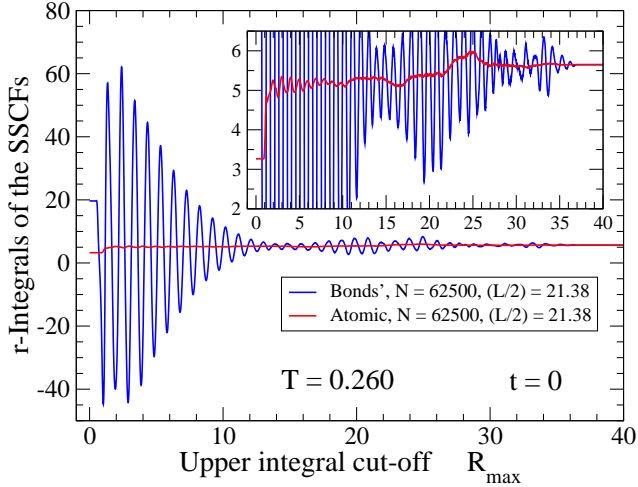}
\caption{The main figure and the inset show on different $y$ scales the 
distance integrals of the ASCF and the BSCF for the system
of 62500 particles at $T=0.260$.
Note that if the integrals are performed over the whole range of distances
then the results of the integrations should be exactly the same if 
the same structures were used for the calculations of the CFs.
This behavior can be observed in the inset.
}\label{fig:int-sscf-atbs}
\end{center}
\end{figure}

As discussed before, the integrals over the distance of the stress CFs in Fig.\,\ref{fig:sscf-at-bs}(a) and Fig.\,\ref{fig:sscf-at-bs}(b)
should lead to exactly the same value which is the value of the macroscopic stress CF
at time $t=0$. This behavior is demonstrated in Fig.\,\ref{fig:int-sscf-atbs}. 

Note in the inset of Fig.\ref{fig:int-sscf-atbs} that the ASCF, 
after obtaining a nonzero value from the zero-distance
term (\ref{eq:F-auto-01}), also abruptly grows as the first nearest neighbors become included. 
Beyond this distance, the integral of the ASCF does not exhibit significant changes.
The variation of the integral for the cutoff distances larger than $10$ are caused by the
limited statistics of the data.
From the perspective of the BSCF the situation looks differently.
The BSCF also obtains nonzero value from the zero distance term (\ref{eq:B-auto-01}).
As distance increases further the BSCF oscillates very significantly around the value 
which is approximately close to the average value of the ASCF.
Thus, from the perspective of the BSCF large distances are very relevant for viscosity. 

In our view, the correlation functions 
$\mathcal{F}_{cross}^{xy}(r)$ (\ref{eq:F-cal-01}) 
and  
$\mathcal{B}_{cross}^{xy}(r)$ (\ref{eq:B-cal-01}),
while of the same origin, actually describe quite different
structural aspects. 
Thus, it appears that the BSCF $\mathcal{B}_{cross}^{xy}(r)$ (\ref{eq:B-cal-01})
indeed primarily describes the correlations in the BOO between individual bonds. 
The atomic level stresses, as has been discussed previously \cite{Egami19801,Egami19802,Egami19821,Chen19881},
essentially describe deviations of the local atomic environments
from some average atomic environment in which the local atomic shear stresses are equal to zero.
Correspondingly, the ASCF describes correlations between the deviations of the local atomic environments
from some average local atomic environment.

\subsection{Distribution of the values of the stress products}

In the context of the presented results, in our view, it is important to discuss
the probability distributions (PDs) of the products 
$\left\langle s_{i}^{xy} s_{j}^{xy}\right\rangle_{\Omega}$,
$\left\langle b_{ij}^{xy} b_{kh}^{xy}\right\rangle_{\Omega}$ and how these PDs 
depend on the distance from $i$ to $j$ or from $ij$ to $kh$. 
This issue, in our view, is also relevant to many previously obtained 
results \cite{Hoheisel19881,Woodcock19911,Woodcock20061,Levashov20111,Levashov2013,Levashov20141,Levashov20142,
Levashov20161,Levashov20162,Levashov20172,Lemaitre20151,HarrowellP20161,Voigtmann2016,Lemaitre20171,Lemaitre20181,Matubayasi2018}, 
though it has not been discussed there.

It is known from numerous previous investigations of the macroscopic Green-Kubo stress CF 
at low temperatures that it is relatively difficult to accumulate good statistics for its long-time
tails \cite{Hoheisel19881}. It is also computationally demanding to accumulate good statistics 
for the microscopic CFs discussed in Ref.\cite{Woodcock19911,Woodcock20061,Levashov20111,Levashov2013,Levashov20141,Levashov20142,
Levashov20161,Levashov20162,Levashov20172,Lemaitre20151,HarrowellP20161,Voigtmann2016,Lemaitre20171,Lemaitre20181,Matubayasi2018}, 
as is mentioned in some of these publications.
In our view, the characters of the mentioned above PDs provide the key for understanding this situation. 
Often, when the average value of some variable
is considered this variable has a not too wide Gaussian distribution around the average value. This is absolutely not the case for the PDs
of   $\left\langle s_{i}^{xy} s_{j}^{xy}\right\rangle_{\Omega}$ and
$\left\langle b_{ij}^{xy} b_{kh}^{xy}\right\rangle_{\Omega}$.

\begin{figure}
\begin{center}
\includegraphics[angle=0,width=3.3in]{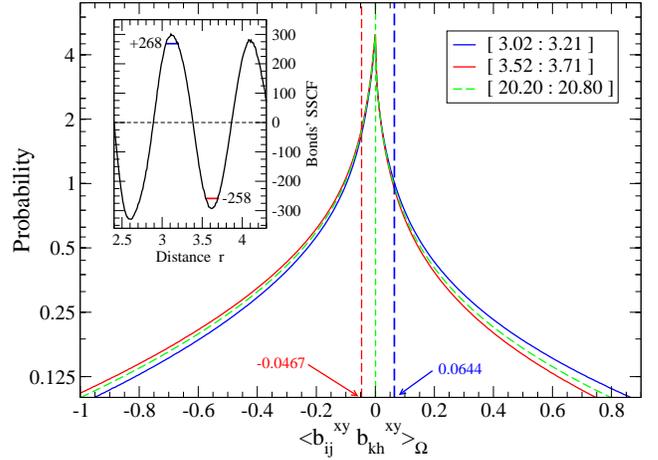}
\caption{The main figure shows the probability distributions (PDs) of the
product $\left\langle b_{ij}^{xy} b_{kh}^{xy}\right\rangle_{\Omega}$
for the $ij$ and $kh$ bonds within three intervals of distances.
The vertical dashed lines show the average values of these
PDs.
The inset shows that the intervals $[3.02:3.21]$ and $[3.52:3.71]$,
used for the calculations of the PDs in the main figure, 
correspond to the positions of the peak and the next minimum in the BSCF. 
The magnitudes of the peak and the minimum are also shown.
}\label{fig:b-ss-pd}
\end{center}
\end{figure}

Figure \ref{fig:b-ss-pd} shows the PDs of the products $\left\langle b_{ij}^{xy} b_{kh}^{xy}\right\rangle_{\Omega}$ 
from the distance intervals $J_1 \equiv [3.02:3.21]$, $J_2\equiv [3.52:3.71]$, 
and $J_3\equiv [20.20:20.80]$. 
The intervals $J_1$ and $J_2$ approximately correspond
to the consecutive maximum and minimum in the BSCF shown in Fig.\,\ref{fig:sscf-at-bs} 
and in the inset of Fig.\,\ref{fig:b-ss-pd}. 
In the inset these intervals are shown as blue and red horizontal line segments.
The PD from the large-distance interval $J_3$ corresponds to the case of nearly uncorrelated bonds. 
See Ref. \cite{prob-distr} for an additional comment.

Note that the differences between all three PDs are small in comparison to the
overall (similar) shapes of the distributions. Yet, these differences are the reason for
the nonzero values of the average BSCF at the considered intervals.

In this paragraph we describe how the values of the BSCF in the inset of Fig.\,\ref{fig:b-ss-pd} 
can be obtained from the average values of the PDs in the main part of Fig.\,\ref{fig:b-ss-pd}.
In our system of $62'500$ particles at $T=0.260$ there are approximately $415'000$ bonds.
The average particles' number density is $\rho_{po} = 0.8$ while the estimated average bonds' number density
is $\rho_{bo}\approx 5.31$. 
In order to evaluate the BSCF per bond, i.e., normalized to the number of bonds, we have to evaluate the value of 
the expression $4\pi r_{ij,kh}^2 \langle\langle  s_{ij}^{xy}s_{ij}^{xy}\rangle_{\Omega}\rangle$, where
$\langle\langle  s_{ij}^{xy}s_{ij}^{xy}\rangle_{\Omega}\rangle$ is the average value 
of a particular bond-stress distribution in the main part of Fig.\ref{fig:b-ss-pd}. 
However, in the inset the BSCF is normalized to the number of particles. 
In order to find the BSCF normalized to the number of particles it is necessary 
to multiply the BSCF normalized to the number of bonds by $(\rho_{bo}/\rho_{po})\approx 6.64$. 
The results of the estimates for the two short-distance intervals are given in Table \ref{table:C2c-tbl}. 
These values are close to the corresponding values of the BSCF in the inset 
of Fig.\ref{fig:b-ss-pd}.
\begin{center}
\begin{table}
\begin{tabular}{| c | c | c | c |} \hline
Interval        & $r_{ij,kh}$ & $\langle\langle  s_{ij}^{xy}s_{ij}^{xy}\rangle_{\Omega}\rangle$ 
& $B_{cross}^{\alpha\beta}(t,r)$  \\\hline
$[3.02 : 3.21]$ & $\sim 3.11$ & $+0.0644$ & $\sim +276.1$ \\\hline
$[3.52 : 3.71]$ & $\sim 3.61$ & $-0.0467$ & $\sim -269.8$ \\\hline
\end{tabular}
\caption{
The values of the parameters that have been used to check the connection between the BSCF normalized 
to the number of particles (\ref{eq:B-cal-01}), i.e.,
$\mathcal{B}_{cross}^{\alpha\beta}(t=0,r)$, and the PDs of 
the correlation products between 
the pairs of bonds at the selected distance intervals.
The connection is given by
$\mathcal{B}_{cross}^{\alpha\beta}(t,r)\sim 4\pi r_{ij,kh}^2 
\langle\langle  s_{ij}^{xy}s_{ij}^{xy}\rangle_{\Omega}\rangle (\rho_{bo}/\rho_{po})$. 
The values of $\mathcal{B}_{cross}^{\alpha\beta}(t=0,r)$ estimated from the PDs 
of the bonds' correlation products are given in the last column. 
}
\label{table:C2c-tbl}
\end{table}  
\end{center}

\begin{figure}
\begin{center}
\includegraphics[angle=0,width=3.3in]{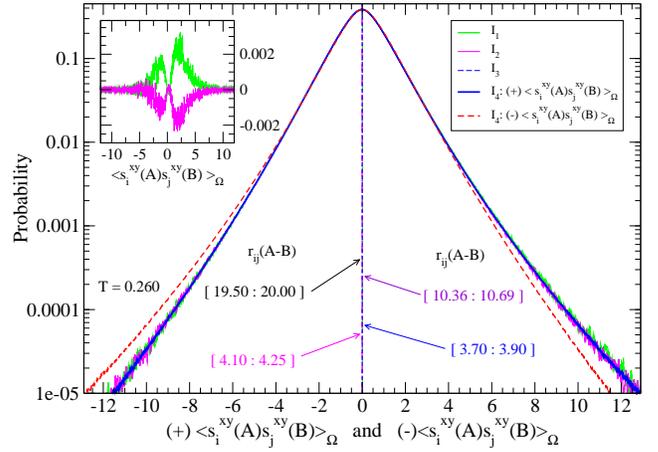}
\caption{The main figure shows the PDs of 
$\langle s_i^{xy}(A)s_j^{xy}(B) \rangle_{\Omega}$ for the ``$AB$" pairs of particles separated by $r_{ij}(AB)$ 
within selected distance intervals:
$I_1 \equiv [3.70:3.90]$,
$I_2 \equiv [4.10:4.25]$,
$I_3 \equiv [10.36:10.69]$,
and 
$I_4 \equiv [19.50:20.00]$.
The two blue curves originating from $I_3$ and $I_4$ cannot be distinguished.
Particles' pairs associated with these large-distance intervals are nearly uncorrelated, 
as follows from Fig.\,\ref{fig:sscf-at-bs}(a). 
The red dashed curve shows the $I_4$ blue curve symmetrically reflected with respect to the zero-abscissa value. 
It shows that the probability distribution curves
are not symmetric even for the well-separated, i.e., independent, pairs of particles.
The green curve originates from the pairs with $r_{ij}(AB) \in I_1$. 
It is also very close to the blue curves.
It follows from the figure that nonzero values of the stress CF
in Fig.\ref{fig:sscf-at-bs}(a) originate from the very small differences between the PDs. 
These differences cannot be observed in the main part of the figure. 
However, these small differences lead to the nonzero values of the ASCF.
With the notation $\xi=\langle s_i^{xy}(A)s_j^{xy}(B) \rangle_{\Omega}$ 
we get $\langle \xi \rangle_{PD} (I_1) = 0.01105$, $\langle \xi \rangle_{PD} (I_2) = -0.00748$,
$\langle \xi \rangle_{PD} (I_3) = 3.41\cdot 10^{-4}$, and $\langle \xi \rangle_{PD} (I_4) = 3.8\cdot 10^{-5}$.
These average values are shown in the figure with the vertical dashed and dot-dashed lines.
These lines cannot be distinguished on the presented scale of the $x$ axis.
However, these invisible differences lead to the nonzero values of the
ASCF shown in Fig.\,\ref{fig:sscf-at-bs}(a).
To make a comparison with the average values of the ASCF in Fig.\,\ref{fig:sscf-at-bs}(a)
it is necessary to multiply the positions of the vertical lines by $4\pi r^2_{ij}(AB)\rho_{p}(AB)$.
Using $r_{ij}(I_1) = 3.8$ and $\rho_{p}(AB) = 0.4$ we get $F_i(t=0,r_{ij}(I_1)) = 0.802$ 
which is in reasonable agreement with the results in Fig.\,\ref{fig:sscf-at-bs}(a).
Assuming that $r_{ij}(I_2) = 4.15$, we get $F_i(t=0,r_{ij}(I_2)) = -0.648$ 
which is also in reasonable agreement with Fig.\,\ref{fig:sscf-at-bs}(a). 
To demonstrate more clearly the small differences between the PDs 
we show in the inset 
the quantities $\left\{\xi\left[P_{1}(\xi)-P_{4}(\xi)\right]\right\}$ 
(upper green curve with positive ordinate values) and
$\left\{\xi\left[P_{2}(\xi)-P_{4}(\xi)\right]\right\}$ 
(lower magenta curve with negative ordinate values),
where $P_{4}(\xi)$ is the PD originating from interval $I_4$, while
$P_{1}(\xi)$ and $P_{2}(\xi)$ are the PDs originating from intervals $I_1$ and $I_2$. 
Since pairs with large separation are essentially uncorrelated 
the integral of $\xi P_{4}(\xi)$ is essentially zero and, as follows, 
the integrals over the green and magenta curves lead
to the average values of the stress correlations associated with $I_1$ and $I_2$.
It is clear that the integral over the green curve is positive, while the integral over the magenta curve is negative.
The main purpose of the figure is to demonstrate that the CFs
shown in Fig.\,\ref{fig:sscf-at-bs}(a) originate from the very small differences 
between the PDs associated with the different distance intervals.  
}\label{fig:a-ss-pd}
\end{center}
\end{figure}

\begin{figure}
\begin{center}
\includegraphics[angle=0,width=3.3in]{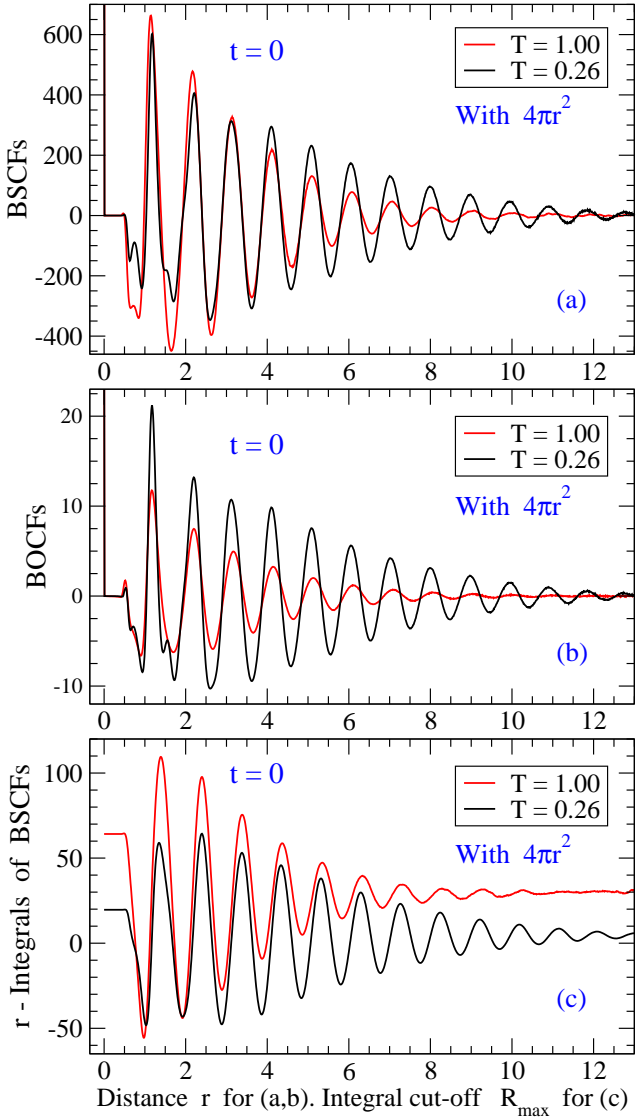}
\caption{Panel (a): The dependencies on the distance of the BSCF at the high and low temperatures.
Note that as temperature decreases the amplitudes of peaks decrease for $r<3\sigma$.
and increase for $r > 3\sigma$. 
Panel (b): The dependencies on the distance of the bond-order CF at the high and low temperatures.
Note that as temperature decreases the amplitudes of peaks increase for all distances, 
in contrast to the situation with the BSCF in panel (a).
Panel (c): The dependencies on the upper cut-off of the integrals over the distance of the BSCFs in panel (a).  
}\label{fig:bcf-diff-T}
\end{center}
\end{figure}

Figure \ref{fig:a-ss-pd} shows the probability distributions 
of the spherically averaged products between atomic stresses, i.e.,
$\left\langle s_{i}^{xy} s_{j}^{xy}\right\rangle_{\Omega}$,
for the ``AB" pairs of particles for the distance intervals 
$I_1 \equiv [3.70:3.90]$, 
$I_2\equiv [4.10:4.25]$, $I_3\equiv [10.36:10.69]$, 
and $I_4 \equiv [19.50:20.0]$.
As follows from Fig.\,\ref{fig:sscf-at-bs}(a),
the $I_1$ interval includes the position of the $fourth$ separate peak
in the ASCF, while $I_2$ includes the position the $fourth$ minimum.
The PDs obtained from the large-distance intervals $I_3$ and $I_4$ 
illustrate the PDs from the nearly uncorrelated particles.

It follows from the figure that the PDs for all selected intervals
are very close to each other. They are essentially indistinguishable.
We note that these distributions are not symmetric with respect to the zero value on 
the horizontal axis even for the pairs of particles at large 
distances, i.e., for the separation intervals $I_3$ and $I_4$. 
This can be seen from the comparison of the
blue curve with its reflection with respect to zero value on $x$ axis.
This reflection is the red-dashed curve.
Despite the similarity of the PDs from the different intervals,
the calculations of the average values 
of $\left\langle s_{i}^{xy} s_{j}^{xy}\right\rangle_{\Omega}$
lead to slightly different results 
which are shown as essentially coinciding vertical dashed lines.
The average values of  
$\left\langle s_{i}^{xy} s_{j}^{xy}\right\rangle_{\Omega}$ from the intervals
$I_1$, $I_2$, $I_3$, and $I_4$ are $0.01105$, $-0.007481$, $3.4116\cdot 10^{-4}$, 
and $3.80417\cdot 10^{-5}$ correspondingly.
These values multiplied by $4 \pi r^2_{I}\rho_p$ lead 
to the approximate magnitudes of the corresponding values 
of the ASCF in Fig.\,\ref{fig:sscf-at-bs}(a).
See the caption of Fig.\,\ref{fig:a-ss-pd} for more details.

The results presented in this section demonstrate that the stress CFs
in Fig.\,\ref{fig:sscf-at-bs}(a,b) originate from rather small differences between the wide probability distributions. 
In our view, the results show that while these average values are related to viscosity via 
the Green-Kubo expression they actually contain rather limited information
about the structure of liquids.

In our view, it is possible to gain some intuitive understanding of the obtained data through a consideration of an example 
usually presented in the context of the fluctuation-dissipation theorem \cite{fluctdissip,Berthier2002}.
Thus, let us think about a very heavy particle that moves through a liquid with a speed which is much smaller than the 
average speed of the light particles comprising the liquid. 
Since the particle is very heavy, we assume that its speed almost does not change.
In addition, let us assume that the size of the heavy particle is not much larger than the size of the liquid's particles. 
In such a situation, the heavy particle experiences forces due to the collisions with the particles of the liquid.
The averaging of these forces over time produces the average viscous force. 
Note, however, that the heavy particle does not
actually experience the viscous force at any particular time; i.e., 
the viscous force is only a convenient way to describe the effect of a very large number of collisions.
It is reasonable to expect that the value of the average viscous force in the described situation should be much smaller than the 
forces arising due to the individual collisions between the particles 
(if the speed of the heavy particle is zero then the average viscous force also should be zero).
Note also that the average viscous force does not correspond to any particular 
collision or any particular type of collisions.
In our view, it is possible to draw a parallel between the average viscous force acting on a particle 
and the average value of the microscopic stress CF at some distance. 
Thus, while the average viscous force results from the averaging over a very large number of ``collision" forces,  
the average value of the stress CF at some distance, $\left\langle s_{i}^{xy} s_{j}^{xy}\right\rangle_{\Omega}$, 
results from the averaging over a very large number of particle pairs' stress products, $s_{i}^{xy} s_{j}^{xy}$.
Then, similarly to the situation with the average viscous force, the average value of the stress correlation function
at some distance, $\left\langle s_{i}^{xy} s_{j}^{xy}\right\rangle_{\Omega}$,
does not correspond to any particular realization of the structural arrangement of the particles. 
Instead, $\left\langle s_{i}^{xy} s_{j}^{xy}\right\rangle_{\Omega}$ 
represents the result of the averaging over a large number of quite different configurations.
For this reason, in our view, one cannot expect to find a structure in the liquid state that actually
would correspond to the average value of the stress correlation function.

In our view, the presented results also elucidate why it is relatively difficult 
to accumulate sufficient statistics for the good quality ASCFs and BSCFs--for this it is 
necessary to produce rather good quality statistics for the wide probability distributions 
of the pair-stress correlations.

\subsection{Evolution of the BSCF with temperature}

Figure \ref{fig:bcf-diff-T}(a) shows how the BSCF depends on distance for
two temperatures (rather high and rather low). 
It is clear that there are no particularly
abrupt qualitative changes in the BSCF on cooling. 
However, there is an interesting change in the behavior of the BSCF with the increase of distance. 
Note that as temperature decreases the amplitudes of peaks
for distances $r<3\sigma$ decrease, while for $r > 3\sigma$ increase.
Previously, in discussing Fig.\,\ref{fig:sscf-at-bs}(b), 
it has been demonstrated that oscillations in the BSCF reflect
mostly the BOO, but not the correlations in the bonds'
tensions.
As temperature decreases there are fewer and fewer strongly compressed bonds in the systems. 
In order to demonstrate the effects associated with the bonds' tensions we show in 
Fig.\,\ref{fig:bcf-diff-T}(b) the bond-order CF analogous to the BSCF in Fig.\,\ref{fig:bcf-diff-T}(a); i.e.,
in order to produce panel (b) it was assumed that $b_{ij}=1$ for all bonds.
The comparison of panel (b) with panel (a) suggests that the decrease in the amplitudes of peaks in
the BSCF for $r<3\sigma$ on cooling is associated with the changes in the bonds' tensions. 
It is also clear from the comparison that the increase in the amplitudes of peaks for $r > 3\sigma$
on cooling is caused by the increase in the BOO.

The changes in the BSCF and bond-order CF in Fig.\ref{fig:bcf-diff-T}(a,b) at large distances 
are clearly of interest due to their relation to viscosity. 
Thus, the integrals over all distances of the BSCFs in panel (a) 
are the contribution to viscosity from $t=0$ at the discussed temperatures.
The dependence of such integrals of the two BSCFs in panel (a) 
on the upper limit of the integration is presented in panel (c).
We see that the contribution to viscosity from $t=0$ in 
the high-temperature liquid is larger than in the low-temperature liquid.
However, it is necessary to remember that the BSCF in 
the high-temperature liquid quickly decays with time; i.e.,
the larger value of viscosity in the low-temperature liquid is due 
to the slow decay of the BSCF associated with the slow $\alpha$ relaxation.

In our view, the most puzzling point with respect to the average values of the stress CFs concerns the results presented in the previous section. 
Thus, due to the large deviations of the particular values
of $\left\langle b_{ij}^{xy} b_{kh}^{xy}\right\rangle_{\Omega}$ 
or 
$\left\langle s_{i}^{xy} s_{j}^{xy}\right\rangle_{\Omega}$
from their average values, it is not quite clear what 
we can learn about the structure or dynamics of the system from 
the considerations of the average CFs.

\subsection{Shear Stress Waves}

\begin{figure}
\begin{center}
\includegraphics[angle=0,width=3.3in]{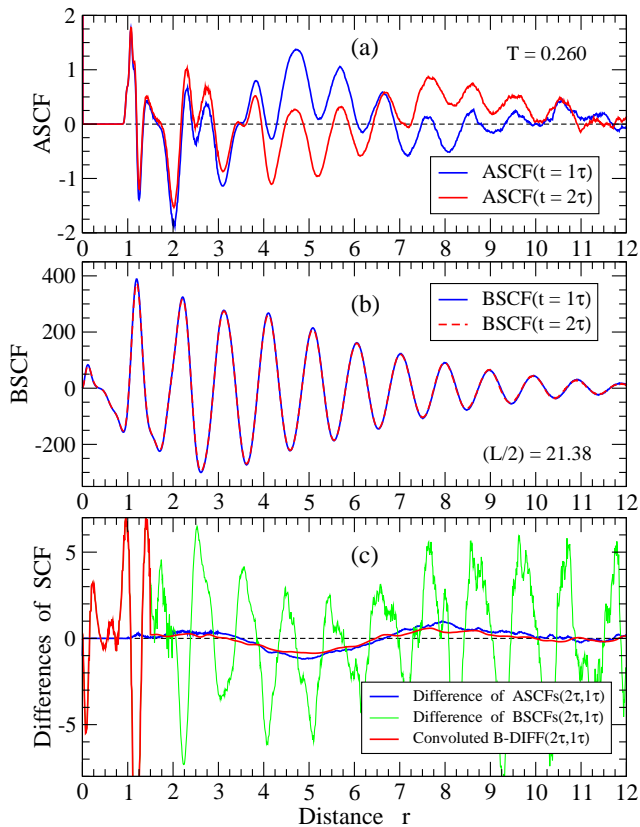}
\caption{Panel (a) shows the ASCFs at $T=0.260$ for 
the times $t= 1\tau$ and $t=2\tau$. 
Note that the blue curve ($t=1\tau$) exhibits a broad shift to the negative intensities, 
obscured by the oscillations, in the interval of distance $1.5<(r/\sigma)<3.5$, while it also exhibits 
a broad shift to the positive intensities in the interval $3.5<(r/\sigma)<6.5$. 
The red curve, corresponding to $t=2\tau$, exhibits a broad negative intensity in the interval
$3.5<(r/\sigma)<6.5$ and a broad positive intensity in the interval $6.5<(r/\sigma)<10.0$.
Thus, as time increases, the regions of negative and positive intensities shift to larger distances.
As has been discussed in Refs.\cite{Levashov20111,Levashov2013,Levashov20141,Levashov20142,Levashov20172} 
this shift of positive and negative intensities with the increase 
of time reflects the propagation of shear stress waves in the system.
Panel (b) shows the BSCF calculated on the same configurations that have been used to produce panel (a). 
In contrast to panel (a), the curves corresponding to $t=1\tau$ and $t=2\tau$ are indistinguishable in panel (b).
Thus, the shear stress waves cannot be observed in the BSCFs themselves. Note, however, the very different scales on the $y$ axis
in panels (a) and (b). Thus, it may not be surprising that shear stress waves with the amplitudes shown in panel (a) cannot be seen in panel (b).
The shear stress waves can, however, be extracted from the BSCF curves with the treatment of the data presented in panel (c).
The blue curve in panel (c) shows the difference between the red and blue curves shown in panel (a). 
The green curve in panel (c) shows the difference between the dashed red and the blue curves presented in panel (b).
The red curve in panel (c) shows the result of the convolution of the green curve with the Gaussian of width $\sigma = 1$.
We see that the blue and red curves in panel (c) are very close to each other and that both of them exhibit a broad minimum and a broad peak
which can be (should be) interpreted as the signatures of the propagating shear waves.
}\label{fig:shear-waves}
\end{center}
\end{figure}

In our previous studies of the ASCFs on two different system of particles we observed
in the data very pronounced features that have been interpreted as propagating compression
and shear waves \cite{Levashov20111,Levashov2013,Levashov20141,Levashov20142,
Levashov20161,Levashov20162,Levashov20172}. See also Ref. \cite{Matubayasi2018}. 
This interpretation initially has been based on the ``speeds of propagation"
of the observed features which are close to the expectable speeds of the longitudinal and transverse waves. 
Later, the ASCFs have been analytically derived for a simple model of a crystal with phonons \cite{Levashov20142}. 
The ASCFs calculated in this model exhibited features which also should be interpreted as propagating
shear and compression waves; for the considered model there is no any other alternative.
The similarity of the features in the ASCFs obtained within the crystal model analytically with the features
observed in the ASCFs obtained numerically on the model liquids also provides support for the 
interpretation of the observed features as signatures of the propagating compression and shear waves.
Finally, in the previous works, we have demonstrated the non-trivial 
effects associated with the periodic boundary conditions \cite{Levashov2013,Levashov20172}. 
The independence of the macroscopic value of viscosity 
from the sizes of finite systems with the periodic boundary conditions also can be explained through the shear stress waves that leave 
and reenter the simulation box because of the periodic boundary conditions.
The major result of those studies is the demonstration that approximately half of the value of viscosity 
is associated with the structural rearrangements, while the other half is associated with 
the character of propagation of the shear stress waves.

Thus, in view of our previous works and also the works of others, it is of interest to address 
the existence of the features in the BSCF that can be interpreted as propagating shear waves. 
This issue is addressed in Fig.\,\ref{fig:shear-waves}. 
As follows from the figure and its caption, the shear stress waves cannot be observed 
in the BSCF as easily as they can be observed in the ASCF. 
However, panel (c) of the figure demonstrates that 
shear stress waves are also present in the BSCF and they can be 
revealed through a couple of simple manipulations with the BSCFs.

The reason because of which the shear stress waves are non-observable in the BSCF directly is related to the values of the tension
of the bonds and changes in these values due to the propagating waves. 
These changes are much smaller than the tensions of the bonds. 
It is also reasonable to assume that the propagating waves do not alter significantly the directions of 
the majority of the bonds. Under these conditions one indeed can expect that the propagating 
shear waves may not be observable in the BSCF directly. 
The situation with the ASCF is different because all pair forces acting on a selected particle mostly compensate each other. 
The noncompensated remaining force is much smaller than the average bond's tension associated with the nearest neighbors. 
This remaining force might be comparable to the changes in the remaining force due to the propagating shear waves. 
This is likely to be the reason because of which the shear stress waves are directly observable in the ASCF.
 
In earlier articles on the mode-coupling theory (MCT) 
it has been demonstrated that, in order to properly describe
the decay of the transverse current CF, it is necessary to introduce the coupling between 
the density fluctuations and the transverse current correlation 
function \cite{Kirkpatrick19841,Kirkpatrick19861,Goetze19981,
Balucani19871,Balucani19881,Balucani19901,Marchetti19921,Das20021,Egorov20081,Fuchs20171}. 
In a recent article, in order to describe the decay of the macroscopic shear stress CF, 
the coupling between the short-time dynamics and the long-time hydrodynamic modes related 
to the transverse current correlation function also has been considered \cite{Fuchs20171}.
In our view, the features that we observe in the ASCF and which we interpret as shear 
stress waves should correspond to the transverse current CFs discussed within the MCT. 
Finally, recently the behavior of the microscopic shear stress correlation 
function has been addressed theoretically \cite{Fuchs20181,Semenov20181}.
In out view, the results presented in \cite{Semenov20181} support our previous interpretations
of the features that we interpreted as shear and longitudinal stress waves.

\section{Conclusions \label{sec:conclusion}}

In this paper, we addressed the atomic scale nature of the Green-Kubo stress correlation function for 
viscosity using the pair-interactions between the particles as the elementary units of stress. 
Previously, the atomic level stresses have been used as the elementary units of stress by us and other authors \cite{Woodcock19911,Woodcock20061,Levashov20111,Levashov2013,Levashov20141,Levashov20142,
Levashov20161,Levashov20162,Levashov20172,Voigtmann2016,Matubayasi2018}. 
In a different approach the space correlations in the coarse-grained stress fields
also have been considered \cite{Lemaitre20151,Lemaitre20171,Lemaitre20181}.

The major purpose of the reported research was to investigate whether the bound stress correlation functions (BSCFs) 
can provide additional information and new insights with respect to the data already obtained using the atomic stress correlation functions (ASCFs). 
Besides, the considerations of the BSCF allow one to draw a direct parallel and make certain comparisons with the 
results obtained previously within the bond-orientational order (BOO) approach.

The obtained results show that the long-range structural correlations are more pronounced in the BSCFs, 
while the dynamical correlations are better expressed in the ASCFs. 
The long-range bond-orientational order CF considered in the reported work is closely related to the BSCF originating from
the atomic scale Green-Kubo expression for viscosity. The considered long-range bond-order CF is related to the $l=2$ spherical harmonics
and is different from the long-range bond-order CFs previously discussed within the BOO approach. 
The characteristic feature of the bond-order CF related to viscosity is that mutual orientations of the pairs of bonds are relevant 
to it, while the orientations of the bonds with respect to the direction from one bond to another turn out to be irrelevant.
It is of interest to notice that while the considerations of the short-range BOO
are abundant in the literature there appear to be only several publications that address the behavior 
of the long-range bond-order CFs as functions of distance \cite{Steinhardt19811,Steinhardt19831,Grest1991,Tomida19951,Chen19881}. 

The considered BSCF, ASCF, and bond-order CF describe the averaged values of the correlation products (for a given value of the distance, $r$). 
For the selected distances, we considered the probability distributions of the correlation products whose averaging leads to the averaged CFs. 
The results show that the individual realizations of the correlation products can deviate very significantly from their averaged values. 
This result shows, in our view, that the developing long-range correlations that we observe in the
averaged values of the BSCF, ASCF, and the BOO CF are actually so small, after all, that it is not clear
how they can be relevant for the dynamic slowdown in liquids on supercooling. 
On the other hand, according to the Green-Kubo expression, the integrals of these average CFs
over the distance, for every instant in time, is the contribution to viscosity from this instant. 
The very slowly decaying (in time) nonzero values of the distance-integrals of these correlations lead 
to the very large values for viscosity of supercooled liquids.

We also demonstrated that the shear stress waves, previously observed in the ASCFs, 
cannot be observed in the BSCFs directly, but can be easily extracted from the BSCFs.

\section{Acknowledgments}

We are grateful to V.K. Malinovsky, V.N. Novikov, and N.V. Surovtsev for the discussion 
of the obtained results prior to the submission of the manuscript. 
This work was supported by Russian Science Foundation (RSF) Grant No. 18-12-00438. 
The numerical calculations have been performed at Supercomputing Center of Novosibirsk State University.

\appendix

\section{Geometrical correlations behind the expression (\ref{eq:boo-Wl}) for $W_l$ \label{apx:idea}}

For convenience we reproduce here expression (\ref{eq:boo-Wl}):
\begin{eqnarray}
W_{l} \equiv  \sum_{\substack{m_1,m_2,m_3 \\ m_1+m_2+m_3=0}} \left(\Gj{l}{l}{l}{m_1}{m_2}{m_3}\right) \overline{Q}_{l m_{1}}\overline{Q}_{l m_{2}}\overline{Q}_{l m_{3}},\;\;\;\;\;\;
\label{eq:boo-Wl-2}
\end{eqnarray}

In expressions (\ref{eq:boo-Wl},\ref{eq:boo-Wl-2}) every BOO parameter 
for the considered group of bonds, i.e., $\overline{Q}_{lm}$, 
is a sum of the spherical harmonics of the same type associated with the bonds in the group. 
Thus, the product of the three BOO parameters for the group can be decomposed into 
the sum of the contributions from the triplets of bonds.

The left-hand side of (\ref{eq:boo-Wl-2}), by construction, 
does not depend on the orientation of the observation coordinate frame.
Thus, the averaging of the left-hand side of (\ref{eq:boo-Wl-2}) over 
the directions of the observation coordinate frame is just the value 
of the left-hand side in a particular observation coordinate frame.
On the other hand, this value should be equal to the value of the right-hand side 
also averaged over the directions of the observation coordinate frame. 
Thus, as follows from the previous paragraph, the spherical averaging of the right-hand 
side consists of the spherically averaged contributions associated with the triplets of bonds.
In order to get an insight into the geometry associated with the triplets of bonds and influencing  
the value of the CF $W_l$ we consider, as an example, the contribution associated with the
product $Y_{20}(\theta_1,\phi_1)Y_{20}(\theta_2,\phi_2)Y_{20}(\theta_3,\phi_3)$, 
where the angles $\theta_n$ and $\phi_n$ characterize the orientations of the three bonds.
Since $Y_{20}(\theta,\phi) \propto (3\cos^2(\theta)-1)$ (no dependence on $\phi$), 
the spherical averaging of the product of the three spherical harmonics $Y_{20}(\theta,\phi)$
involves the averaging of $\cos^2(\theta_n)$ and the products: $\cos^2(\theta_n)\cos^2(\theta_m)$,
$\cos^2(\theta_1)\cos^2(\theta_2)\cos^2(\theta_3)$.

We averaged these functions of the angles over the directions of the observation coordinate frame 
using the same method that has been used in Ref.\,\cite{Levashov20162} in order to derive 
the expression (\ref{eq:Gs}) of this paper. We performed the necessary analytical 
calculations with the wxMaxima computer program \cite{wxMaxima}. 
Here we provide the final answers without giving more details on 
the procedure described previously (see Appendix A in Ref.\cite{Levashov20162}).
With the notation 
$\hat{r}_n=\left[\cos(\phi_n)\sin(\theta_n),\sin(\phi_n)\sin(\theta_n),\cos(\theta_n)\right]$, the results are the following:
\begin{eqnarray}
\left< \left[\cos(\theta_1)\right]^2\right>_{\Omega} = \frac{1}{3},\;\;\;\;\;\;
\label{eq:cos2-01}
\end{eqnarray}
\begin{eqnarray}
\left< \left[\cos(\theta_1)\cos(\theta_2)\right]^2 \right>_{\Omega} = \frac{1}{15}+\frac{2}{15}\left(\hat{r}_1\hat{r}_2\right)^2,\;\;\;\;\;\;
\label{eq:cos2cos2-01}
\end{eqnarray}

\begin{eqnarray}
&&\left< \left[\cos(\theta_1)\cos(\theta_2)\cos(\theta_3)\right]^2 \right>_{\Omega} = \label{eq:cos2cos2cos2-01}\\
&&-\frac{1}{35}\nonumber + \left(\hat{r}_1 \cdot \left[\hat{r}_2 \times \hat{r}_3\right]\right)^2\nonumber\\
&&+\frac{2}{35}\left[\left(\hat{r}_1\hat{r}_2\right)^2+\left(\hat{r}_1\hat{r}_3\right)^2+\left(\hat{r}_2\hat{r}_3\right)^2\right].\;\;\;\;\;\;\nonumber
\end{eqnarray}



\begin{thebibliography}{10}
\bibitem{prob-distr}
It is also possible to obtain the PD 
of $\left\langle b_{ij}^{xy} b_{kh}^{xy}\right\rangle_{\Omega}$ under the assumption
that there are no correlations in the bonds' orientations and bonds' tensions from the bonds' 
tension exponential PDs presented in Fig.\,\ref{fig:force-distribution}. 
However, it appears that the deviations of the bonds' tension PDs
from the model exponential functions lead to the PD for the BSCF
that is noticeably different from the ``uncorrelated" PD obtained in simulations.
For these reason this model PD is not shown, even thought it is close enough
to the MD results to verify the correctness of the calculations.
\end{thebibliography}
\end{document}